\documentclass[aps,prb,twocolumn]{revtex4}

\pdfoutput=1

\usepackage{epsfig}
\usepackage{amssymb}
\usepackage{amsmath}
\usepackage{graphicx}
\usepackage{subfigure}
\usepackage{textcomp}
\usepackage{color}
\usepackage{amsfonts}
\usepackage{bbold}
\usepackage{dsfont}
\usepackage{epsfig}

\begin{document}

\title{The magnetism of wurtzite CoO nanoclusters}

\author{Ruairi Hanafin, Thomas Archer and Stefano Sanvito}
\affiliation{School of Physics and CRANN, Trinity College, Dublin 2, Ireland}
\date{\today}

\begin{abstract}
The possibility that the apparent room temperature ferromagnetism, often measured in Co-doped ZnO, is due to
uncompensated spins at the surface of wurtzite CoO nanoclusters is investigated by means of a combination of density 
functional theory and Monte Carlo simulations. We find that the critical temperature extracted from the specific heat 
systematically drops as the cluster size is reduced, regardless of the particular cluster shape. Furthermore the presence 
of defects, in the form of missing magnetic sites, further reduces $T_\mathrm{C}$. This suggests that
even a spinodal decomposed phase is unlikely to sustain room temperature ferromagnetism in ZnO:Co.
\end{abstract}
\maketitle

In recent years the search for ferromagnetism in insulating oxides doped with small quantities of transition metals has 
become a topic generating much debate in the literature. Taking ZnO:Co as the proto-typical example for this class of 
materials, many experimental groups have reported room temperature ferromagnetism \cite{Venky,RTFM,gamelin}, 
whereas several other have failed to find any such evidence \cite{Dinia,Joy}. Notably, growth conditions, sample 
morphology and spatial Co distribution play a crucial r\^ole in determining the magnetic properties. In particular, there is 
now an emerging view that samples with high structural quality and uniform Co distribution do not result in long-range 
ferromagnetism at high temperature, and that the magnetism may be related to structural \cite{scott} or point 
defects \cite{das}.

Given the unsettled experimental landscape it should not surprise that a number of interesting and competing 
theoretical models have been proposed. In general explanations involving standard mechanisms for the 
magnetic interaction appear problematic.
Schemes leading to short range magnetic coupling such as super-exchange need a Co concentration, [Co], exceeding the 
percolation threshold. This is around 20~\% for an interaction extending to the nearest neighbour sites of an {\it fcc} lattice, 
and it is about a factor of 5 larger than the typical experimental concentrations. Similarly carrier mediated mechanisms are 
not sustained by experimental evidence, which shows both paramagnetism in presence of abundant free carriers 
\cite{scott} and ferromagnetism deep in the insulating region of the phase diagram \cite{RTFM}. More generally carrier 
concentration and mobility have an apparent little correlation to the magnetic properties \cite{refresh}.
Furthermore carrier mediated mechanisms are difficult to validate on a solid theoretical ground by using first principles 
calculations, since the empty Co $d$ levels are usually erroneously predicted too shallow at the edge of the ZnO 
conduction band \cite{das2}.

Thus one has to look for more complex mechanisms for the magnetic interaction. Amongst these, the donor impurity 
band exchange model (DIBE) \cite{Coeynmat} has enjoyed considerable popularity in the experimental 
community. According to the DIBE the magnetic interaction among Co$^{2+}$ ions is mediated by donors, whose
charge density is localized over large hydrogenic orbitals, so that the relevant percolation threshold becomes 
that of the donors and not that of the Co. 
Unfortunately the model still fails at the quantitative ground, since room temperature ferromagnetism needs 
prohibitively large exchange coupling between the donors and the Co \cite{Coeynmat,meJMMM}. A second, recently 
proposed scheme, is the two-species model \cite{das, Rh1} in which Co-oxygen vacancy pairs (CoV) act as a 
second magnetic center in addition to Co$^{2+}$. Interestingly CoV can interact magnetically strongly up to the
third nearest neighbours, where the percolation threshold drops to 7~\%. Although this is still too high to justify a 
long-range ferromagnetic order, it suggests that room temperature ferromagnetism can be achieved in samples 
where Co ions are non-uniformly distributed. 

The extreme limit of Co segregation is represented by the formation of some secondary Co-based phase. Usually
metallic Co is excluded by X-ray data \cite{Tom12}. Unfortunately all other compounds in the Zn-Co-O phase
diagram (CoO, Co$_2$O$_3$, Co$_3$O$_4$, ZnCo$_2$O$_4$) 
are either non-magnetic or antiferromagnetic with low N\'eel temperatures,
so that there is no obvious candidate material to form ferromagnetic clusters in ZnO. Still, recently Dietl et al. 
suggested that the apparent room temperature ferromagnetism in ZnO:Co may be due to uncompensated 
spins at the surface of CoO nanoclusters maintaing the ZnO wurtzite (WZ) lattice structure \cite{Dietlcluster}. Indeed
both WZ and zincblende (ZB) CoO were synthesized in the bulk \cite{risbud52,risbud53,risbud54}, but little
magnetic characterization exists. Unfortunately calculations based on density functional theory (DFT) and Monte 
Carlo (MC) methods are discouraging \cite{tom}. In fact all the possible CoO polymorphs,
including WZ and ZB, do not display a ferromagnetic order but instead they are characterized by a general spin 
frustration. As a result, the critical temperatures, $T_\mathrm{C}$, extracted from the specific heat are well below 
room temperature. Notably such a frustration was recently confirmed experimentally \cite{CoeyCoO} for the
WZ phase.

However, even if one can exclude magnetism in the bulk, the question of whether or not uncompensated spins
can order at the surface of a nanocluster remains open. In particular this is an intriguing question since
the dominant exchange parameter extracted from DFT for wurtzite CoO, the one giving frustration, is large. One
can then speculate that reducing frustration (as it happens on a surface) can help in enhancing $T_\mathrm{C}$. 
Such an open question is investigated here, where we use MC simulations for extracting the thermodynamical 
properties of WZ CoO nanoparticles of different shapes and dimensions. 

\section{Foundation of the model and Computational Details}

We perform MC calculations for a classical Heisenberg Hamiltonian of the form
\begin{equation}
H=-\frac{1}{2} \sum_{i,j} {J}_{ij} \vec{{S}_{i}}\cdot \vec{{S}_{j}} + 
\sum_{i}D\left( \vec{S_{i}}\cdot \hat{n} \right)^{2}\:,
\label{eq1}
\end{equation}
where $\vec{S}_i$ is a classical spin located at site $i$ ($|\vec{S}_i|=3/2$ for Co$^{2+}$) and $J_{ij}$ is the exchange parameter
between spins at sites $i$ and $j$. The second term in (\ref{eq1}) describes the hard-axis easy-plane crystal anisotropy,
$\hat{n}$ is a unit vector along the WZ $c$-axis and the zero-field splitting for Co$^{2+}$ taken from electron paramagnetic 
resonance~\cite{tom21} is $D=2.76 $~{cm}$^{-1}$.

The values for the various exchange parameters, $J$, for bulk WZ CoO have been calculated previously \cite{tom} from 
DFT using the LDA+$U$ extension of the local density approximation (LDA). The procedure used was
to fit the total energies of a number of reference DFT calculations for different magnetic supercells to the Hamiltonian of 
equation (\ref{eq1}). For WZ CoO four different $J$'s are enough to describe the dominant magnetic interaction, with 
an error over the reference DFT calculations smaller than 1~meV/Co. We have then found: $J_1=6.1$~meV, 
$J_2=-36.7$~meV,  $J_3=-0.2$~meV, $J_4=-5.2$~meV, where the index $n$ in $J_n$ refers to the neighbour degree
(i.e. ``1'' is for first near neighbours Co$^{2+}$). The dominant interaction, $J_2$, is thus for first nearest neighbours
in the $\{001\}$ plane (second nearest neighbours overall) and it is antiferromagnetic. This is the origin of the frustration
and of $T_\mathrm{C}$'s below room temperature for the bulk. The same parameters are used for the calculations of
finite particles presented here. 

The ground state of a given particle is found by the simulated annealing method, since frustration introduces many low energy
configurations differing considerably from the ground state but with minor energy differences from it. Such an energy landscape
clearly makes the conjugate gradient and the steepest descent schemes ineffective. The temperature dependence is then investigated 
with Monte Carlo simulations. We use the standard Metropolis algorithm where the acceptance probability of a new state is 
100~\%, if the new state has an energy lower than that of the old one, and it is given by the Boltzmann factor otherwise. For each
different magnetic cluster we equilibrate the system at a given temperature and then extract thermodynamical quantities by sampling
over several millions MC steps. In particular we extract $T_\mathrm{C}$ from the peak in the specific heat, $C$. This 
becomes necessary since an obvious order parameter is difficult to find for these highly frustrated clusters. 

Such a computational scheme is not free of uncertainty. Firstly, the reference DFT calculations are dependent on the 
specific choice of exchange and correlation function used. LDA+$U$ is certainly suitable for CoO and our parameterization 
of the $U-J$ parameter is based on total energy considerations, i.e. on fitting the structural properties and not the band-structure 
\cite{tom}. Secondly the Heisenberg model used contains only $J$'s extending over a limited range and includes only pairwise
interaction. The first approximation appears acceptable given the good quality of the fit to DFT, while the second one is
more difficult to assess. Nevertheless we have used rocksalt CoO to estimate the error and find that typically our $T_\mathrm{C}$'s
for the bulk are underestimated by about 30~\% \cite{tom}. Considering that the Co valence in rocksalt and WZ CoO is the
same, we speculate that the same error found for the rocksalt phase can be transfered to WZ. 

Still the uncertainty of using bulk parameters for finite clusters simulations  remains. However, it is important
to bare in mind that here we do not consider free-standing CoO nanoparticles, but instead CoO clusters embedded into
a ZnO matrix. This means that the local chemical coordination of each Co atom (i.e. the fourfold coordination to O)
in the cluster is identical to that of bulk CoO. Thus the finite size affects only the magnetic coordination and one can 
safely use bulk parameters for clusters. This, of course, would be inadequate in the case of free surfaces or grain 
boundaries for which a new parameterization is needed.

\section{Results}

If present, it is likely that CoO nanoclusters form in ZnO either during the growth process or during post-growth treatment. 
DFT calculations in fact confirm that there is an energy gain when moving Co ions to nearest neighbouring positions \cite{das},
so that clustering is highly probable. This can possibly be tuned by growth parameters and eventually the presence of
additional dopants \cite{dietlnatmat}. Still, even assuming that ZnO:Co during the growth has enough kinetic energy to form
large CoO clusters, it is hard to predict whether such clusters should have a preferential shape. Indeed simulations
of spinodal decomposition \cite{KY} seem to suggest the formation of highly Co-rich regions
without any particular geometrical structure. This is then a different situation from that of free-standing nanoparticles, where
the presence of free surfaces drives the particle geometry. For this reason we have looked at three different
particle shapes, namely spherical, cylindrical and perforated spherical. The last ones are spherical particles where 
a fraction of the Co sites is randomly removed.

\subsubsection{Spherical Nanoparticles}

We consider spherical particles first. These are constructed by simply removing all the atoms beyond a sphere of a 
give radius, $R$. In doing so the position of the center of the sphere influences the atomic details of the external 
surface so that different particles with the same radius can be made. We find that all the particle properties
are relatively sensitive to the position of the particle center (chosen within the WZ unit cell) for small particles, but 
they become progressively center-independent as the radius gets larger. This is expected since larger surfaces 
allow all the thermodynamical quantities to self-average. For this reason we perform averages over the position of 
the center only for the smaller particles investigated. 

\begin{figure}[!h]
\epsfxsize=7cm
\centerline{\epsffile{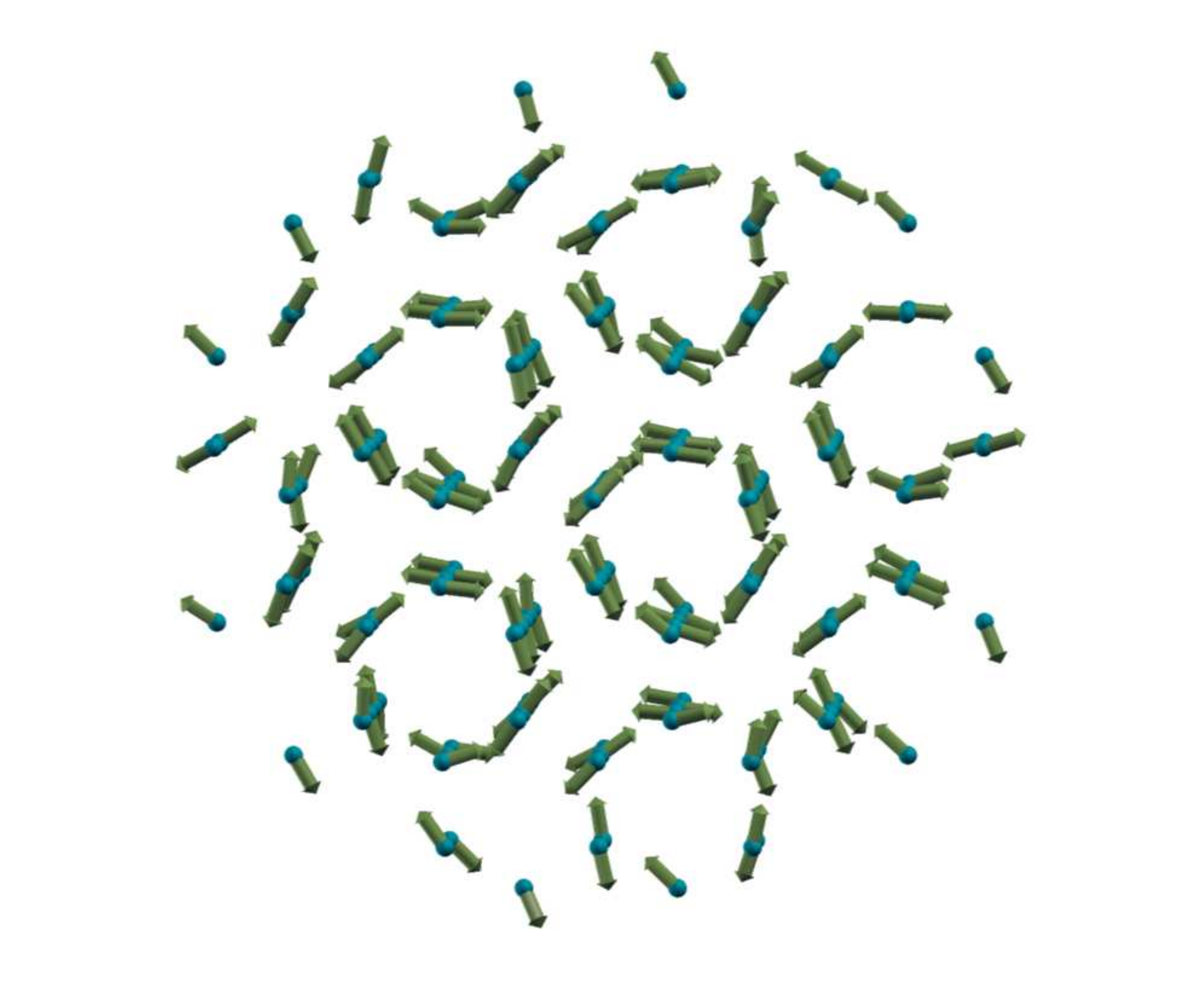}} 
\centerline{\epsffile{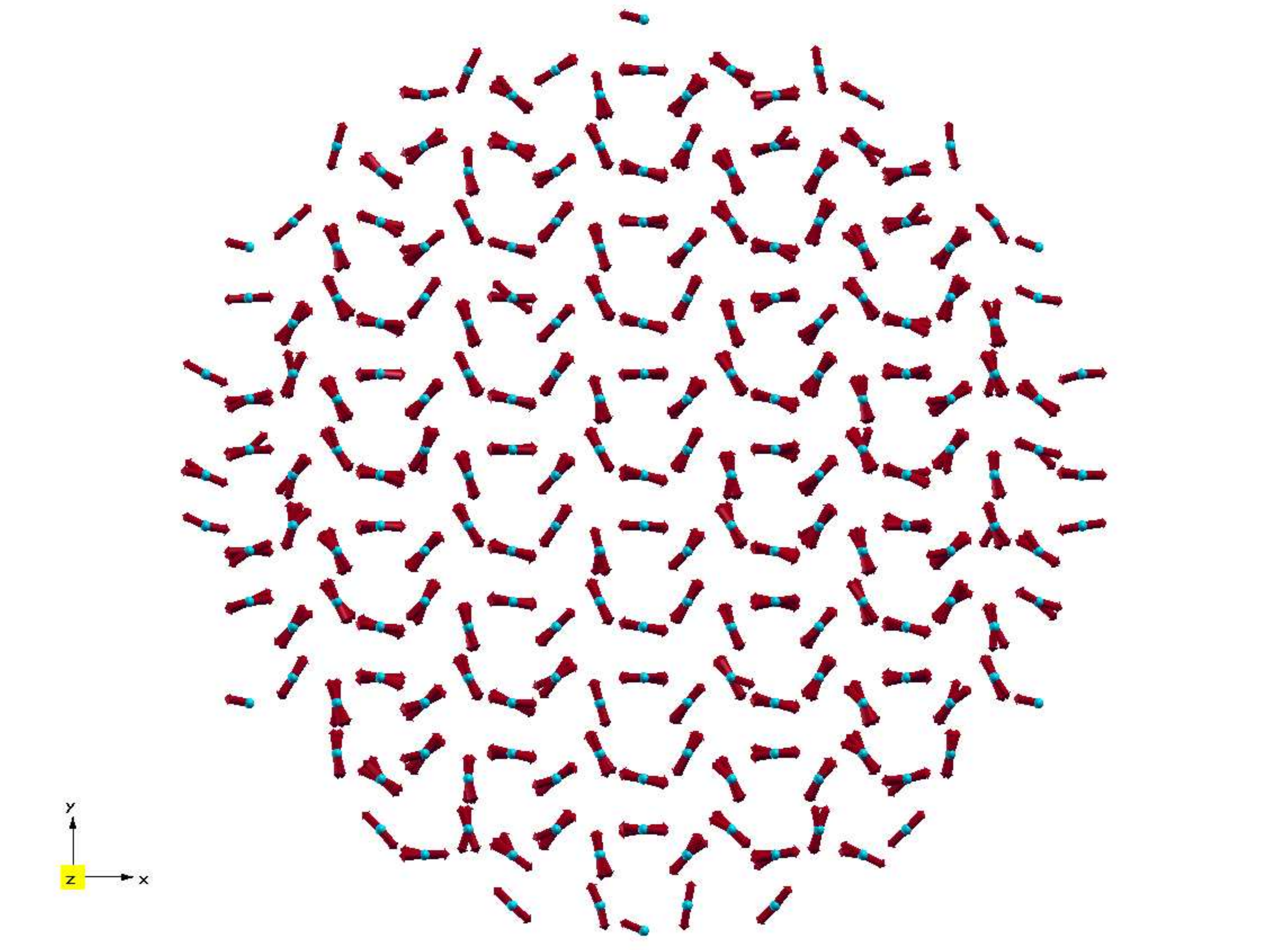}} 
\caption{[Color on line] Ground state spin configuration of spherical particles of different size: $12$~\AA\ (top) and 
$18$~\AA\ (bottom). Note the large degree of compensation in the inner core of the particles and the presence of 
non-compensated spins at their surfaces.}
\label{Fig1}
\end{figure}
We start our analysis by presenting the ground state spin configuration as calculated with simulated annealing. 
This is shown in figure~\ref{Fig1} respectively for a small ($R=12$~\AA) and a large ($R=18$~\AA) particle.
In the figure the particles are oriented with the WZ $c$-axis pointing perpendicularly to the page, so that the $a$-$b$
planes are visualized. 
The figure displays a clear spin compensation of the particle inner core. As for the case of bulk WZ CoO, the strong
first nearest neighbour antiferromagnetic constant in the $a$-$b$ plane, $J_2$, drives the frustration. As a result 
the spins in the $a$-$b$ plane align at 120$^o$ with respect to each other and the net moment per plane vanishes. 
At the surface the frustration is lifted by translation symmetry breaking and a net uncompensated moment, $\mu$, emerges. 
Its direction and intensity depends on the particle size, and for small particles on the details of the surface geometry.
In any case the moment is always rather small and it is then difficult to visualize by simply looking at figure~\ref{Fig1}.

In table~\ref{Tab1} we list such a ground state uncompensated magnetic moment (per Co atom) for spherical 
particles of different sizes.
\begin{table}
\begin{center}
\begin{tabular}{cccccc} \hline\hline
 $R$ (\AA)   & $V$ (\AA$^3$) & $S$ (\AA$^2$) & $N_\mathrm{Co}$  & $\mu$/Co ($\mu_\mathrm{B}$) &  $T_\mathrm{C}$   \\
	\hline
6   & 940 & 452 & 52  & $< 0.225$& 81  \\
7   & 1436 & 615 & 64 & $< 0.225$ & 100  \\
8   & 2143 & 803 & 94 & $< 0.18$ & 107\\
10   & 4186 & 1256 & 174 & $< 0.18$ & 126  \\
12  & 7234 & 1808 & 324 & $< 0.09$ & 136\\
17  & 20569 & 3629 & 920 &   $< 0.045 $ & 163\\
19  & 28716 & 4534 & 1300 &  $< 0.03 $   & 169\\
22  & 44580 & 6079 & 2016 &  $< 0.015$   & 176\\
29  & 102109 & 10562 & 4634 &   $< 0.0015$  & 187\\	\hline\hline
\end{tabular}
\caption{Table listing the uncompensated magnetic moment per Co in the ground state, $\mu$, and the
critical temperature, $T_\mathrm{C}$, extracted from the specific heat, for spherical particles of different radius.
We also report the number of Co atoms contained in the particle, $N_\mathrm{Co}$, and both particle volume, $V$,
and area of the surface, $S$.}
\label{Tab1}
\end{center}
\end{table}
In the table for each radius we report the upper value obtained over a number of different geometrical realizations 
of the particle. In general the uncompensated magnetic moments are rather small, so that the cluster model itself is 
at best capable of explaining only weak magnetism in ZnO:Co. The table also allows us to extract an upper limit
for the magnetic moment at room temperature due to WZ CoO particles. 
In fact, by assuming that magnetization reversal is driven by coherent rotation we estimate that particles with 
radii larger than 15~\AA\ (containing about 800 Co atoms) are large enough to be superparamagnetically blocked 
at room temperature. Their ground state (at $T=0$) magnetic moment is calculated between 0.1~$\mu_\mathrm{B}$/Co 
and 0.05~$\mu_\mathrm{B}$/Co (Table~\ref{Tab1}). Thus, if magnetism originates from CoO clusters with uncompensated 
spins, then a ferromagnetic signal at room temperature cannot be associated to magnetic moments in excess of  
$0.1~\mu_\mathrm{B}$/Co. In fact particles with larger $\mu$ have a radius smaller than 15~\AA, they are not be 
superparamagnetically blocked at room temperature, and hence they cannot contribute to the ferromagnetic signal. 
However, even for the possibility of weak ferromagnetism to be sustained one has to demonstrate that the 
small magnetic moments survive at room temperature. 

\begin{figure}[h]
\epsfxsize=8cm
\centerline{\epsffile{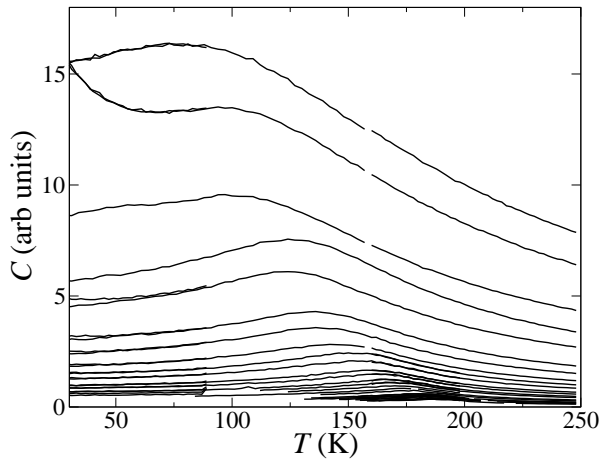}}
\caption[The specific heat, $C$, as a function of temperature for finite clusters of wurtzite CoO.]
{The specific heat, $C$, as a function of temperature for finite clusters of wurtzite CoO. The curves are for
particles of different radii, ranging from 6~\AA\ (top) to 30~\AA\ (bottom), in increment of 1~\AA\ }
\label{Fig2}
\end{figure}

Unfortunately for these frustrated system the magnetization is not a good order parameter and no sublattices 
magnetizations can be identified. Moreover both the uncompensated moment and the particle susceptibility 
turn out to be rather noisy quantities, so that no $T_\mathrm{C}$ can be extracted from them. We then
calculate $T_\mathrm{C}$ from the analysis of the specific heat, $C$, as a function of temperature. This is
shown in figure~\ref{Fig2} for a number of nanoparticles of different radii. The upper curve corresponds to 
$R=6$~\AA\ and the lower to $R=30$~\AA\ with the curves in between corresponding to radii incrementing 
by 1~\AA. 
From the curves we can clearly identify a peak that becomes more diffuse as the particle size grows. Nevertheless
we can still assign to the peak position the $T_\mathrm{C}$ of each particle. These are also reported in Tab.~\ref{Tab1}.
Interestingly the $T_\mathrm{C}$ calculated for bulk WZ CoO with the same parameterization used here is 
only 160~K \cite{tom}, i.e. it corresponds to a particle with a radius of 16~\AA.

\begin{figure}
\epsfxsize=8cm
\centerline{\epsffile{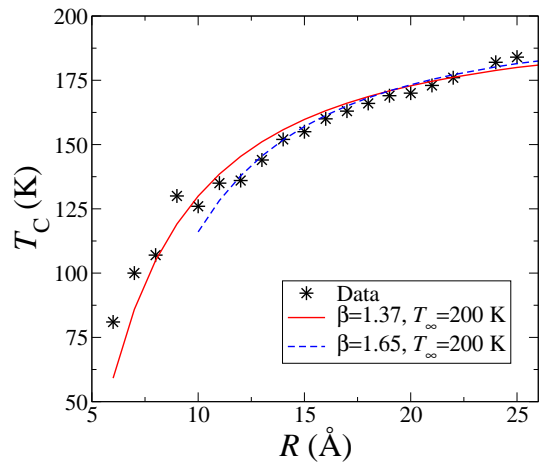}} 
\caption[$T{_c}$ as extracted from the specific heat curves for clusters of different radius.]{[Color on line] 
Critical temperature as a function of the particle radius $R$ ($*$ symbols). The lines are fits to finite size scaling theory 
[Eq.~(\ref{eq2})]. The solid red line corresponds to the fit obtained for $R>6$~\AA\ , while the dashed blue 
curve is for $R>10$~\AA.}
\label{Fig3}
\end{figure}

Finite size scaling theory for the Heisenberg model and ferromagnetic interaction\cite{FSST1,FSST2} predicts a variation of the 
critical temperature with particle radius of the form 
\begin{equation}
\frac{T_\infty-T_\mathrm{C}(R)}{T_\infty}=\left(\frac{R}{R_0}\right)^{-\beta}\:,
\label{eq2}
\end{equation}
where $R_0$ and $T_\infty$ are respectively the correlation radius and the critical temperature for the 
bulk. This cannot be applied directly to our results since $T_\infty$ is smaller than $T_\mathrm{C}(R)$
for large clusters, i.e. the left-hand side of equation (\ref{eq2}) becomes negative for some $R$.
We have then fitted the calculated $T_\mathrm{C}(R)$ to the equation~(\ref{eq2}), by assuming a 
different $T_\infty$. The results are reported in figure~\ref{Fig3}. In the curve we show two fits obtained by
taking $T_\infty=200$~K and performing the fit respectively over the range $R>6$~\AA\ (solid red curve) and 
$R>10$~\AA\ (dashed blue curve). As expected the fit improves when particles with small radii
are excluded, since the scaling low is strictly valid in the vicinity of $T_\infty$. In any case we
clearly observe that a rather good fit can be obtained for critical exponents in the range of that 
expected for the Heisenberg model ($\beta\sim1.42$), when $T_\infty$ is about 200~K. For these 
values we find a correlation radius of the order of 6~\AA.

We are at this time uncertain of why $T_\infty$ calculated from MC for an infinite system (in practice it is calculated for a 
finite system with periodic boundary conditions and sufficiently large cells) differs from that extrapolated from finite
size scaling. In general $C(T)$ for finite particles is much more diffuse than that calculated by using periodic boundary 
conditions, so that the assignment of $T_\mathrm{C}$ may be affected by some errors. However, such an uncertainty 
is smaller than the difference between the two temperatures ($\sim$40~K), so that an alternative explanation 
is needed. We believe that the surface and the core of the particles contribute in a substantial different way to the
$T_\mathrm{C}$ so that scaling theory cannot be applied directly. In fact, depending on the magnetic coupling
at the surface compared to that of the bulk, it was already demonstrated that $T_\mathrm{C}$ can either increase 
or decrease as a function of the particle size \cite{FT,Wel,Rus}. This means that depending on the details of the 
magnetic interaction $T_\mathrm{C}$ as a function of the particle size can approach $T_\infty$ asymptotically either 
from below or from above. In the present case of frustrated interaction it appears that the bulk value is approached in
a non-monotonic fashion so that there exits spherical particles with a $T_\mathrm{C}$ larger than
that of the bulk. This aspect will be further investigated in the next section. 

In any case we find that spherical particles, despite for large radii present a $T_\mathrm{C}$ larger than that of the
bulk, still fall quite short from being magnetic at room temperature. Furthermore the particles with the largest magnetic
moment are those showing the lowest $T_\mathrm{C}$, which makes us concluding that spherical wurtzite Co particles
cannot be at the origin of the room temperature magnetism in ZnO:Co.

\subsubsection{Cylindrical Nanoparticles}

\begin{figure}
\epsfxsize=8cm
\centerline{\epsffile{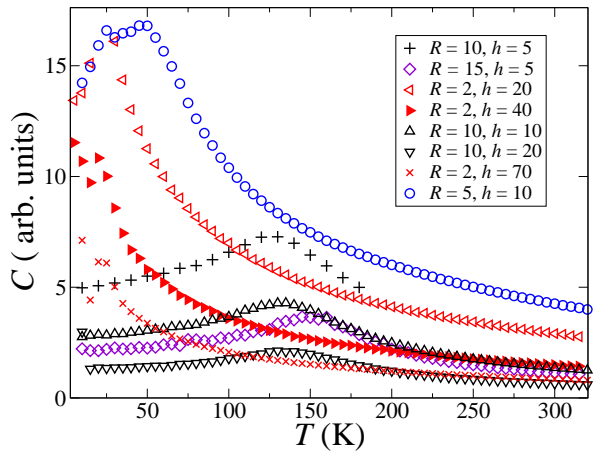}} 
\caption{[Color on line]. Specific heat as a function of temperature for a representative sample of cylindrical 
nanoparticles with different radii and lengths.}
\label{Fig4}
\end{figure}

In order to further investigate the relative importance of the surface and bulk contributions to the magnetism we consider
cylindrical nanoclusters, constructed with the cylinder axis oriented along the WZ $c$-axis. By varying both 
the length and radius of such a particle the surface area may be considerably altered whilst the total volume and 
hence the total number of atoms in the cluster remains constant. In figure~\ref{Fig4} we present the specific 
heat for a number of cylinders of different dimensions, where we can still clearly observe the presence of a peak.
As in the case of spherical particles we associated the critical temperature to the peak position. 

These are presented next in table~II, where we list also the particle volume and the area of the surface.
The most relevant feature emerging from the table is the rather sharp dependance of $T_\mathrm{C}$ on the particle
radius and its insensitivity over its length. This results in the interesting finding that particles presenting the same volume 
but different aspect ratios can display rather different $T_\mathrm{C}$'s. For example the volume of a particle with 
$R=5$~\AA\ and $h=30$~\AA\ is about 30\% larger than that of a particle with $R=10$~\AA\ and $h=5$~\AA. Nonetheless
its $T_\mathrm{C}$ is a factor 3 smaller (44~K against 122~K). 

In addition we find that the critical temperature of cylindrical nanoparticles becomes almost independent from the 
length of the cylinder beyond a certain critical length, $h_\mathrm{C}$. For instance $h_\mathrm{C}$ is 2~\AA, 5~\AA, 
10~\AA\ and 20~\AA\ respectively for $R$ being 2~\AA, 5~\AA, 10~\AA\ and 15~\AA. Thus it is tempting to propose the
empirical relation $h_\mathrm{C}\sim R$. As a consequence for particles with $R/h<1$ there is only one relevant dimension, $R$. 
In fact for long cylinders ($R\ll h$) the surface to volume ratio remains constant at $2/R$, so that the magnetic energy density 
does not change with the cylinder length. This seems to be at the origin of the saturation of $T_\mathrm{C}$ with
$h$.

Interestingly for $R=h$ the area of the surface of the cylinder and that of the sphere become identical, while the volumes 
follow the relation $V_\mathrm{sphere}=4/3V_\mathrm{cylinder}$. In general we expect that the $T_\mathrm{C}$
of a nanoparticle has both a surface and a volume contribution. In the case $R=h$ the surface contribution is 
identical for spheres and cylinders, while the volumetric contribution is expected to be larger for the cylinders, since these have
a larger volume. Thus we expect that the $T_\mathrm{C}$ of a sphere of radius $R$ is lower than that of a cylinder of 
radius $R$ and $h=R$, as indeed confirmed by comparing the tables I and II. 

\begin{table*}[htb]
\begin{center}
\begin{ruledtabular}
\begin{tabular}{c|ccccccc}
$R|h$ &  2    &   5  & 10  &  20  &  30  &  40  &  60     \\
	\hline
2   & {\bf 16}$_{(25, 50)}$ &{\bf 18}$_{(63, 88)}$ & {\bf 21}$_{(125, 150)}$  & {\bf 20}$_{(251, 276)}$ & {\bf 20}$_{(377, 402)}$ & {\bf 21}$_{(502, 527)}$ & {\bf 23}$_{(754, 779)}$\\
5   & {\bf 26}$_{(157, 219)}$ & {\bf 46}$_{(392, 314)}$ & {\bf 46}$_{(785, 471)}$ & & {\bf 44}$_{(2355, 1099)}$ & & \\
10   & {\bf 102}$_{(628, 753)}$ & {\bf 122}$_{(1570, 942)}$ & {\bf 135}$_{(3140, 1256)}$ & {\bf 135}$_{(6280, 1884)}$ & {\bf 135}$_{(9420, 2512)}$ & {\bf 138}$_{(12560, 3140)}$ & {\bf 138}$_{(18840, 4396)}$  \\
15   & {\bf 122}$_{(1413, 1601)}$ & {\bf 152}$_{(3532, 1884)}$ & {\bf 160}$_{(7065, 2355)}$ & {\bf 167}$_{(14130, 3297)}$ & {\bf 170}$_{(21195, 4239)}$ & {\bf 171}$_{(28260, 5181)}$ &  \\
25   & {\bf 128}$_{(3925, 4239)}$ & {\bf 158}$_{(9812, 4710)}$ & {\bf 171}$_{(19625, 5495)}$ & {\bf 181}$_{(39250, 7065)}$  & & & \\
30   & {\bf 130}$_{(5652, 6029)}$ & {\bf 165}$_{(14130, 6594)}$ & {\bf 183}$_{(28260, 7536)}$ & {\bf 191}$_{(56520, 9420)}$ & & & \\
\end{tabular}
\label{Tab2}
\caption{The calculated $T_\mathrm{C}$ for a number cylindrical nanoparticles with different radius, $R$, and length, $h$.
The radius is along the vertical axis and the length is along the horizontal one (both in \AA). In brackets we report respectively 
the cylinder volume (\AA$^3$) and surface area (\AA$^2$).}
\end{ruledtabular}
\end{center}
\end{table*}
Finally we take a look at the magnetization, finding that this is always small, typically $< 10^{-3}\mu_\mathrm{B}/\mathrm{Co}$, 
and varies little with either changes in the particle dimensions or with the temperature. In summary, from our analysis it does not 
appear that cylindrical nanoparticles, similarly the spherical ones, are supportive of room temperature ferromagnetism.

\subsubsection{Perforated Spherical Nanoparticles}

Since empty Co sites in a CoO nanoparticle may serve to lift the spin frustration by eliminating a fraction of the
neighbours of each Co atom, it is worth analyzing the dependence of $T_\mathrm{C}$ over the Co concentration. 
This essentially corresponds to studying a random ZnO:Co alloy with large Co doping. In general we expect 
that, as long as the Co concentration exceeds the percolation threshold for the wurtzite lattice (19~\% for nearest
neighbour interaction), a magnetic order will be found. The question remaining is whether or not magnetization 
and $T_\mathrm{C}$ will increase with respect to their values for stoichiometric CoO when the Co concentration 
is reduced. 

The nanoparticles investigated here are constructed by taking one of the previously made spherical nanocrystals and then 
removing a chosen fraction of magnetic ions at randomly chosen sites. We consider particles with a radius of
$19~$\AA, which are large enough to show a substantial $T_\mathrm{C}$, but not enough for the thermodynamical
properties to saturate at their bulk values, i.e. the surface of the particles still makes a non-negligible contribution.
Furthermore we find that particles with smaller radii are too sensitive to Co vacancies, to the extent that in general 
at Co concentrations below 90\% a $T_\mathrm{C}$ is not easily located. Note that, although we name Co vacancy a
site where the Co ion is removed, the parameterization for the exchange interaction remains the one
calculated for the perfectly crystalline CoO phase \cite{tom}, so that the atom removal has only geometrical
effects.

\begin{figure}
\epsfxsize=8cm
\centerline{\epsffile{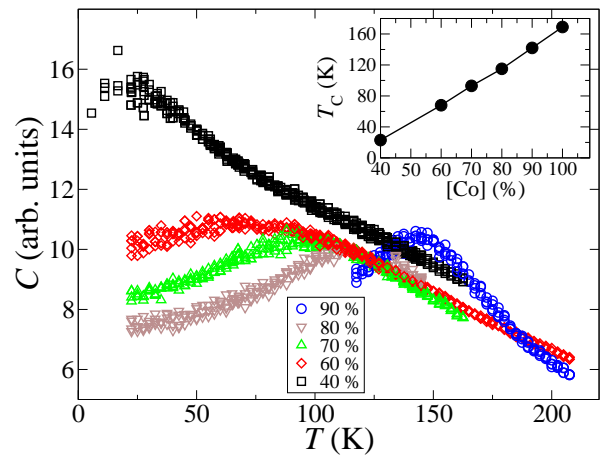}} 
\caption[Specific heat curves against temperature for a number of perforated nanospheres of radius $19$\AA\ with 
varying concentration of atomic vacancies.]
{[Color on line] Specific heat against temperature curves for a perforated nanosphere of radius 19~\AA\ and various
Co concentrations, [Co]. We can clearly observe that by reducing the Co content the peak in $C(T)$ moves to lower temperature
and gets broader. In the inset we show $T_\mathrm{C}$ as a function of [Co] as extracted from the specific heat.}
\label{Fig5}
\end{figure}

Our results are presented in Fig.~\ref{Fig5} where we show the specific heat as a function of temperature 
for different Co concentrations, [Co]. In the same figure we also report the extracted $T_\mathrm{C}$'s as a 
function of [Co]. The most notable feature is that the peak in $C(T)$ shifts towards lower temperatures and
gets broader as Co atoms are removed. This means that the presence of defects reduces the critical
temperature, so that lifting locally the frustration does not help in improving the magnetic properties. 
We find that $T_\mathrm{C}$ decreases linearly with the fraction of vacant sites and extrapolates to 
$T_\mathrm{C}=0$ for [Co]=$30$\%, which is larger than the nearest neighbour percolation threshold. 
This, however, should not be surprising since deviation from the linear dependence are expected at
low [Co]. Interestingly such a linear dependance of $T_\mathrm{C}$ on the defect concentration has been
reported previously in both experimental and theoretical studies for Ni-Cu alloys \cite{Rus,Kud}

Finally we observe that none of the structures investigated present magnetizations greater than 
0.003~$\mu_\mathrm{B}$/Co above 100~K. We then conclude that the breaking of frustration by means of defects
appears to have little impact on the magnetization and as a consequence perforated nanoclusters do not appear 
to sustain any room temperature ferromagnetism.

\section{Conclusions}
The focus of this paper was to test the feasibility of the proposal that uncompensated spins on the surface 
of CoO nanoclusters embedded in ZnO are responsible for the measured magnetic properties of ZnO:Co. 
Previously we have demonstrated that bulk wurtzite CoO displays a high degree of frustration so that no net magnetization
is found even at low temperature. Furthermore the $T_\mathrm{C}$ calculated from the peak in the specific heat 
is substantially lower than room temperature. 

The present study has revealed that finite sized clusters are no more promising. Although the frustration is lifted at
the interface, we found that usually the finite size reduces the critical temperature for magnetism with respect to its
bulk value. This happens regardless of the shape of the particle and of the presence of Co empty sites. For some
relatively large spherical and cylindrical nanoparticles we have found a marginal enhancement of $T_\mathrm{C}$
with respect to bulk, for which we speculate on a non-monotonic dependence of $T_\mathrm{C}$ with
particle size. In any case the residual magnetizations originating from the finite size remain extremely small
at any temperature, so that finite particles can hardly be considered at the origin of the claimed ferromagnetism
of ZnO:Co. 

In concluding we wish to remark once again that our model is based on a parameterization of the magnetic
interaction rooted in DFT calculations for bulk CoO. It is indeed possible that such an interaction is substantially 
altered at surfaces, so that much larger exchange constants may be found either at grain boundaries or at free 
surfaces or at the interfaces with the substrate. This can promote residual ferromagnetism. 
Furthermore our present model does not include intrinsic defects, which can both alter the local magnetic
coupling or form complexes with Co which may interact over a long range \cite{das}.

\end{document}